\documentclass[aps,twocolumn,epsfig,showpacs]{revtex4}
\usepackage[normalem]{ulem}
\usepackage{subfigure}
\usepackage{graphicx}
\usepackage{epsfig}
\usepackage{amsmath}
\usepackage{color}
\usepackage{mathtools}

\usepackage{hyphenat}
\let\oldref\ref
\renewcommand{\ref}[1]{(\oldref{#1})}

\begin{document}
\title{Periodically driven DNA: Theory and simulation}
\author{Sanjay Kumar}
\affiliation{Department of Physics, Banaras Hindu University, Varanasi 
221\,005, 
India}
\author{Ravinder Kumar}
\author{Wolfhard Janke}
\affiliation{Institut f\"ur Theoretische Physik, Universit\"at Leipzig, 
Postfach 100\,920, D-04009 Leipzig, Germany}
\begin{abstract}
We propose a generic model of driven DNA under the influence of an oscillatory 
force of amplitude $F$ and  frequency $\nu$ and show the existence of a dynamical 
transition for a chain  of finite length. We find that the area of the hysteresis loop, $A_{\rm loop}$,
scales with the same exponents as observed  in a recent study based on a much more 
detailed model. However, towards the true thermodynamic limit, 
%\sout {we find that} 
the high-frequency scaling regime extends to lower frequencies for larger chain length $L$ and the
system 
has only one scaling ( $A_{\rm loop} \approx \nu^{-1}F^2)$. 
Expansion of an analytical expression for $A_{\rm loop}$ obtained for the 
model system in the low-force regime revealed that there is a new scaling exponent associated with force 
($A_{\rm loop} \approx \nu^{-1}F^{2.5}$), which has been validated by high-precision numerical calculation. 
By a combination of analytical and numerical arguments, we also deduce that for  
large but finite $L$, the exponents are robust and independent of temperature and friction coefficient.
\end{abstract}
\maketitle
Living systems are open systems and hence never in equilibrium. Biological processes, e.g., 
transcription and  replication of nucleic acids, packing of DNA in a capsid,
synthesis and degradation of proteins etc., are driven by different types of
molecular motors {\it in vivo} \cite{albert}. These motors act like a repetitive 
force generator due to the chemeomechanical cycles resulting through the hydrolysis of ATP
\cite{tom,patel,jan, velankar, wang, basu,fili,janovjak}. 
 Surprisingly, application of an oscillatory force remains elusive in single molecule force spectroscopy (SMFS)
experiments. Rather a  constant force or loading rate frequently used in SMFS experiments have enhanced our understanding 
\cite{Liphardt,Collin,Schlierf,kumarphys}, but provided a limited  picture of these processes. For
example, by varying the frequency of the applied force, it is possible to
observe a dynamical transition, where without changing the physiological condition, 
the system may be brought from the zipped or unzipped state to
a new dynamic (hysteretic) state \cite{maren,kapri,kumar,arxiv1,mishra,kapri1}. Thus, the application of oscillatory 
force will open a new domain of observations and provide further insight into these processes, 
which would not be possible in the case of a steady force. 
\begin{figure}[t!] 
	\vspace*{2mm}\includegraphics[scale=.28]{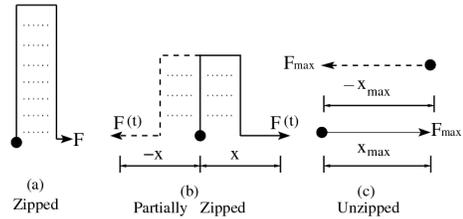}
% vspace NEEDED: Otherwise Fig.~1 partially hides text of the abstract !!!
 \caption{
\label{fig:scheme1}Schematic representations of DNA: (a) zipped, 
(b) partially zipped, and (c) unzipped state. One end is kept fixed (indicated by the
solid circle), while the other end may move in positive (shown by the solid line) 
or negative direction (shown by the dashed line)  depending on the
force direction.}
\end{figure}

\begin{figure*}[!t]
        \centering
        \subfigure{\includegraphics[scale=0.32]{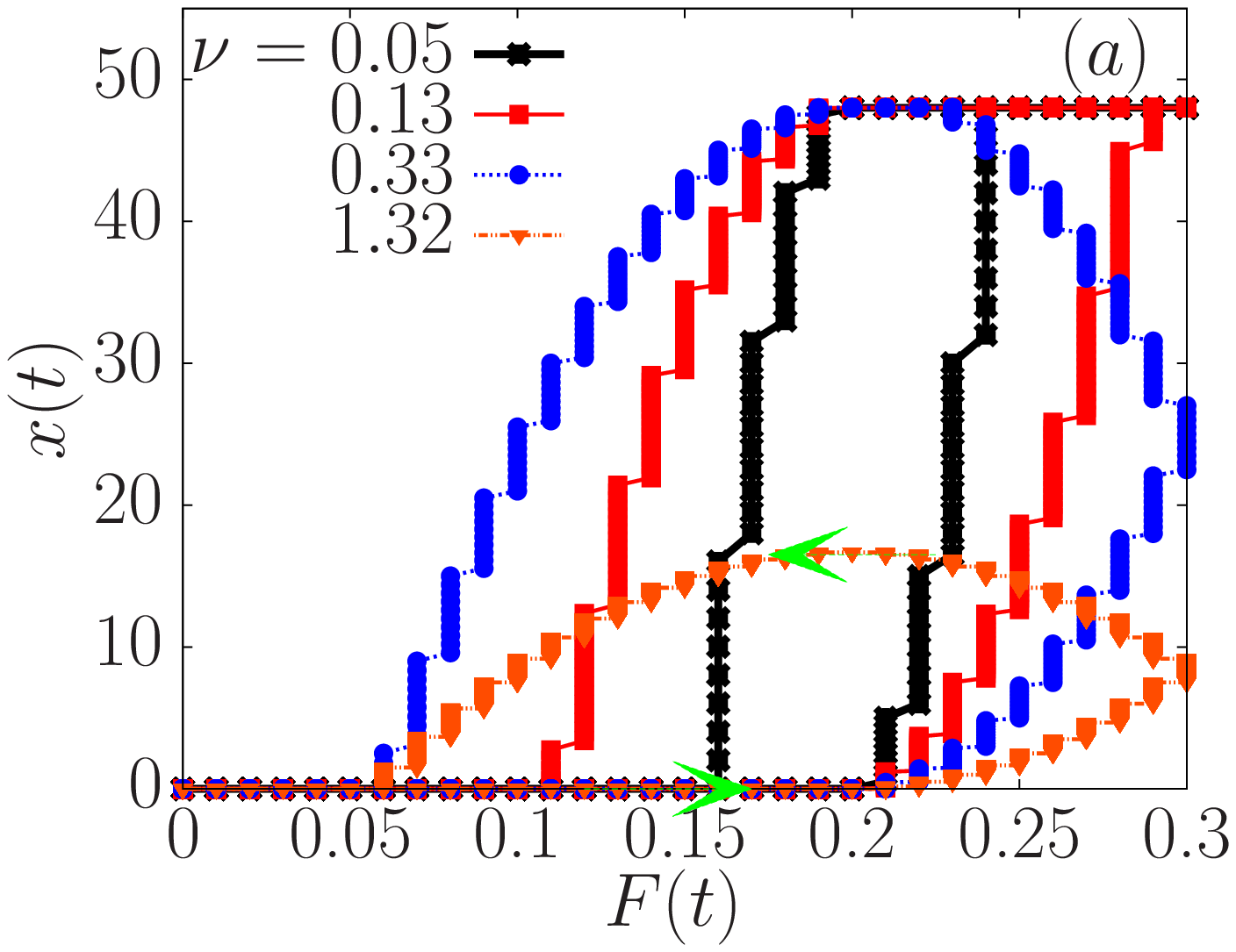} \label{step_area}}
        \subfigure{\includegraphics[scale=0.32]{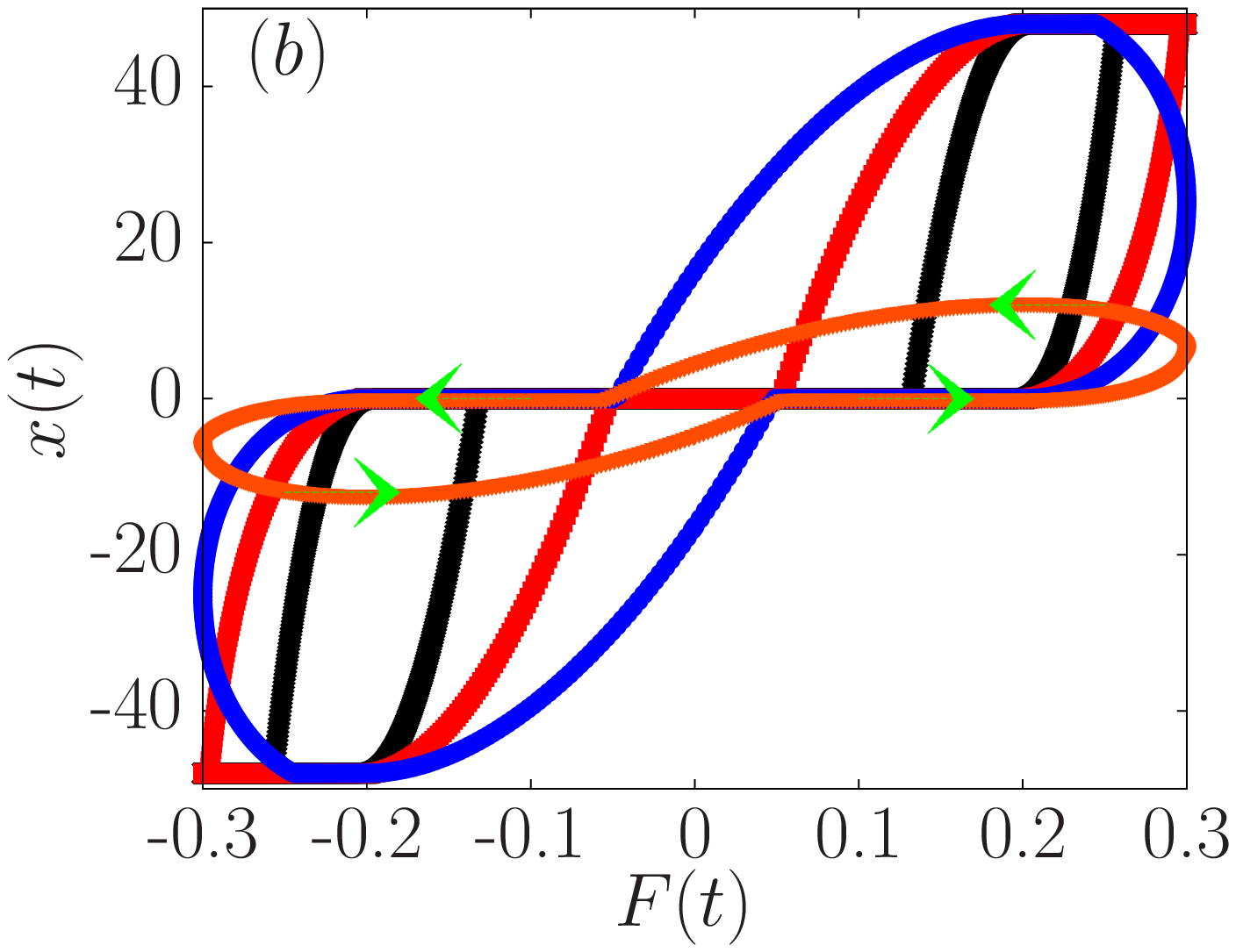} \label{sinf_area}} 
        \subfigure{\includegraphics[scale=0.32]{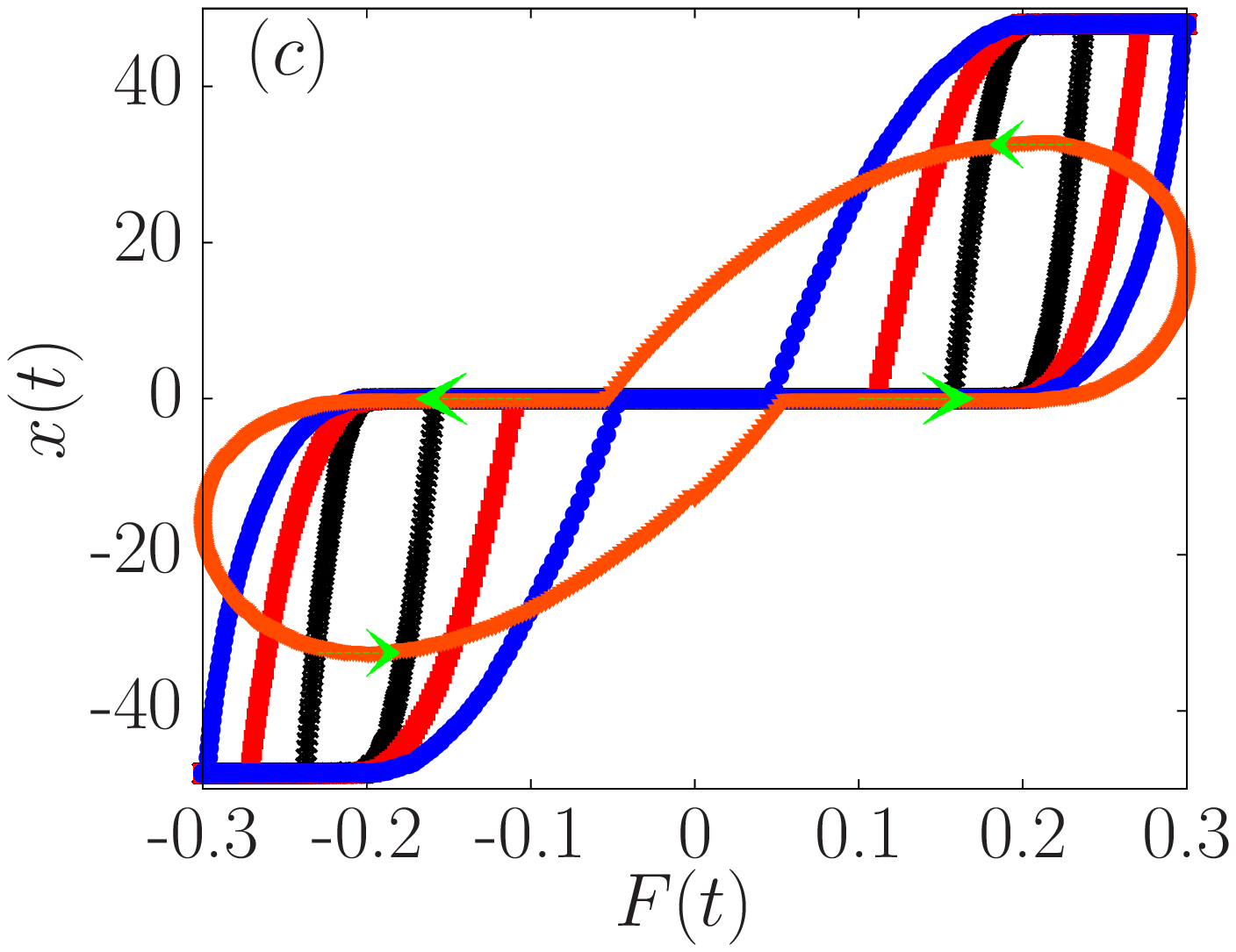}\label{odl_area} }
        \caption{(Color
  online)  Comparative plots of the area of hysteresis loop  for (a) staircase and (b), (c)
sinusoidal force for $F = 0.33$ and chain length $ L = 24$. In (c) the over-damped limit is taken.
For the sinusoidal force, $\nu$ was  taken as $\nu /2$ to compare the positive branch of the
hysteresis loop with the staircase force.  Here $\nu$ is given in units of
$10^{-3}$.
}
\label{fig:area_all_temp}
\end{figure*}
When a DNA chain is driven  by  an oscillatory force, a finite relaxation time produces a 
lag between force and response, and hence produces hysteresis
\cite{maren,kapri,kumar,arxiv1,mishra,kapri1}.
The area of hysteresis loop, $A_{\rm loop}$, under a periodic  force with amplitude $F$ and
frequency $\nu$
was found numerically in a rather detailed model
to scale as $F^{\alpha}\nu^{\beta}$ \cite{kumar}. Here, $\alpha$ and $\beta$ are the characteristic
exponents 
similar to the ones seen in the case of isotropic spin systems \cite{madan1,madan2,dd,bkc}. 
Using Langevin dynamics (LD) simulations for different chain lengths, Mishra {\em et al.\/} \cite{mishra} 
found that these exponents remain independent of solvent quality 
(varying friction coefficient) and interactions involved in the stability of 
bio-molecules (e.g., native interaction for DNA and non-native 
interaction for a polymer globule). Moreover,  they also reported
the dependence of loop area on the length of the chain, which shows
a power-law scaling.  In the low-frequency regime, the area 
of the hysteresis loop per nucleotide $A_{\rm loop}/N$ scales as $F^{0.5} 
\nu^{0.5}N^{0.75}$, where $N$ is the total number of nucleotides. However, scaling arguments suggest
that $A_{\rm loop}/N$
should scale as $N^{0.5}$. In the high-frequency limit, $A_{\rm loop}$ remains 
independent of the chain length with $\alpha =2$ and $\beta = -1$.  Employing Monte Carlo
simulations
on two interacting directed random walks, Kapri \cite{kapri1} observed that in the high-frequency
limit, 
the scaling exponents remain the same, whereas at low frequency, he reported $\alpha= 1$ and $\beta
= 5/4$. 
At this stage, there is no unanimity, thus these discrepancies must be resolved either by 
longer simulations based on the realistic model of DNA or through a minimal
model for which an analytic solution can be derived.

The model and method  adopted in Ref.\ \cite{kumar} 
can describe equilibrium and  non-equilibrium aspects of DNA quite well
\cite{Li,rkm,nath,cieplak,libio}, 
but simulations of longer chain length appear to be computationally challenging.  An analytical
solution of this model is not easy
because of the many degrees of freedom involved. The aim of this work is to successively
reduce  the complexity of the model system and to identify the distinguishing degrees of freedom and
parameters involved and 
thereby to develop a minimal model to understand the underlying mechanism behind the robustness of
these
scaling exponents. 

In this spirit,
we first revisited the mesoscopic  model proposed in Ref.\ \cite{kumar} and performed LD simulations
\cite{Allen, Smith} at different
temperatures ($T = 0.1, 0.08,$ and $0.06$) for a fixed length ($N = 32$). Remarkably, for all these
temperatures,
the values of the exponents remain the same. Thus, one can speculate that these exponents are
insensitive to temperature
and for a better understanding of the dynamics, the system can also be studied at $T = 0$.
In the following, we consider two interacting strings of length $L$ (Fig.~1) to model 
DNA \cite{text0}. One end of the DNA is fixed, and an oscillatory force $F(t)$ is applied on the
other end. 
The total energy of such system can be expressed as 
\begin{equation}
	E=-\epsilon_{\rm bp} N_p + \frac{\epsilon_{\rm bp}}{2a}|x| - F(t)x \, ,
\end{equation}
where $N_p$, $\epsilon_{\rm bp}$ and $|x|$ are the total number of base pairs, the base pairing
interaction and 
the length of the unzipped part of the DNA, respectively. The length of the completely unzipped DNA 
is given by $x_{\rm max} =2L= 2aN_p$, where $a$ $(=1)$ is the distance between two adjacent bases.
%The dynamics of this system at $T = 0$ can be obtained  by using the following equation of motion
Following equation of motion has been used to study the dynamics of the system at $T = 0$
 \cite{Allen,Smith}:

\begin{equation}
m\frac{d^2{x}}{dt^2} = -{\zeta}\frac{d{x}}{dt} - {\epsilon} \frac{{x}}{|x|}+F(t) \, ,
\label{lang}
\end{equation}
\noindent where $\epsilon \equiv \epsilon_{\rm bp}/2$ has been set equal to 0.2. Here, $m$ $(=1)$ and
$\zeta$ $(=0.4)$ 
are the mass of the string and friction coefficient, respectively. 
We used  the fourth-order Runge-Kutta method (RK4) to solve Eq.\ (2) \cite{Allen, Smith} with
time step $dt = 0.01$.  The singularity at $x=0$ was removed by considering $x=0$ as part of one 
of the two non-singular domains, i.e., ${\epsilon} \frac{{x}}{|x|}$ is replaced by $\epsilon$ for
$x \ge 0$ and $-\epsilon$ otherwise. This prescription leads to small oscillations of the 
numerical solution
around $x=0$ and creates an error of the order of $dt^4 =10^{-8}$ \cite{new}.
The system achieved a steady state after about 10 cycles, but we took averages after 100 cycles.

For a staircase like periodic force \cite{kumar}, the force-extension ($F-x$) curves for length
$L=24$ are 
depicted in Fig.~2(a) for different frequencies $\nu$. The qualitative nature of these curves
remains the same as seen at finite temperature ($T=0.1$) in the mesoscopic model.
In order to arrive at an analytic solution, we
choose $F(t) = F\sin(\omega t)$, where $\omega =2\pi\nu$. In Fig.~2(b), we have plotted the $F-x$ 
curves for this sinusoidal force. One can see the existence of hysteresis at different frequencies, 
but due to the sinusoidal nature of force, the extension will also go in the negative direction 
\cite{text1}. In the over-damped limit, the contribution of inertia term on the l.h.s.\ of Eq.\ (2)
is small and can be dropped. We perform Brownian dynamics (BD) simulations to obtain $F-x$ curves
[Fig.~2(c)].
 It is evident from all these plots that the qualitative behavior of the hysteresis does not change.
 
\begin{figure}[t]
        \centering
        \subfigure{\includegraphics[scale=.29]{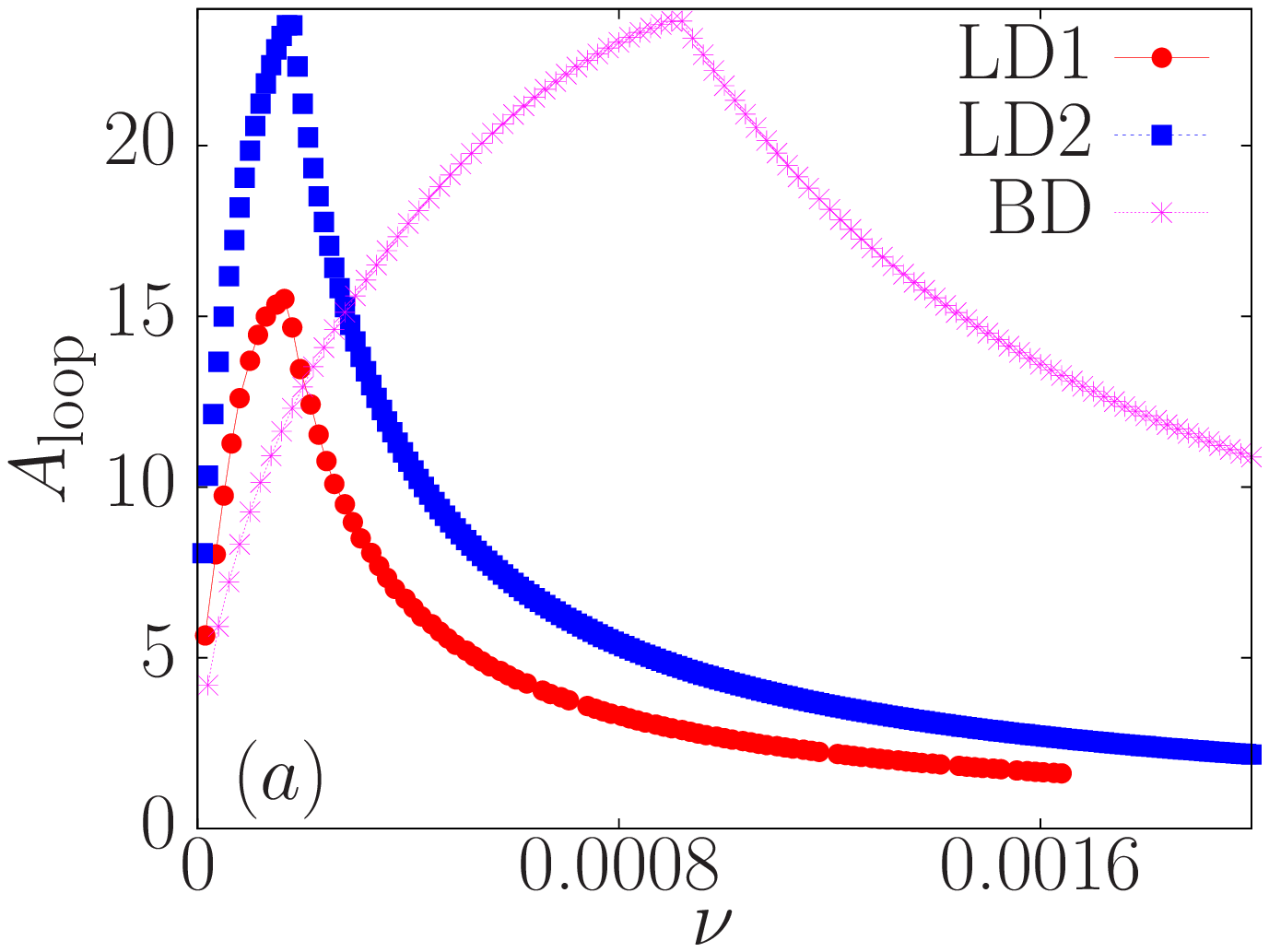}}
	\subfigure{\includegraphics[scale=.29]{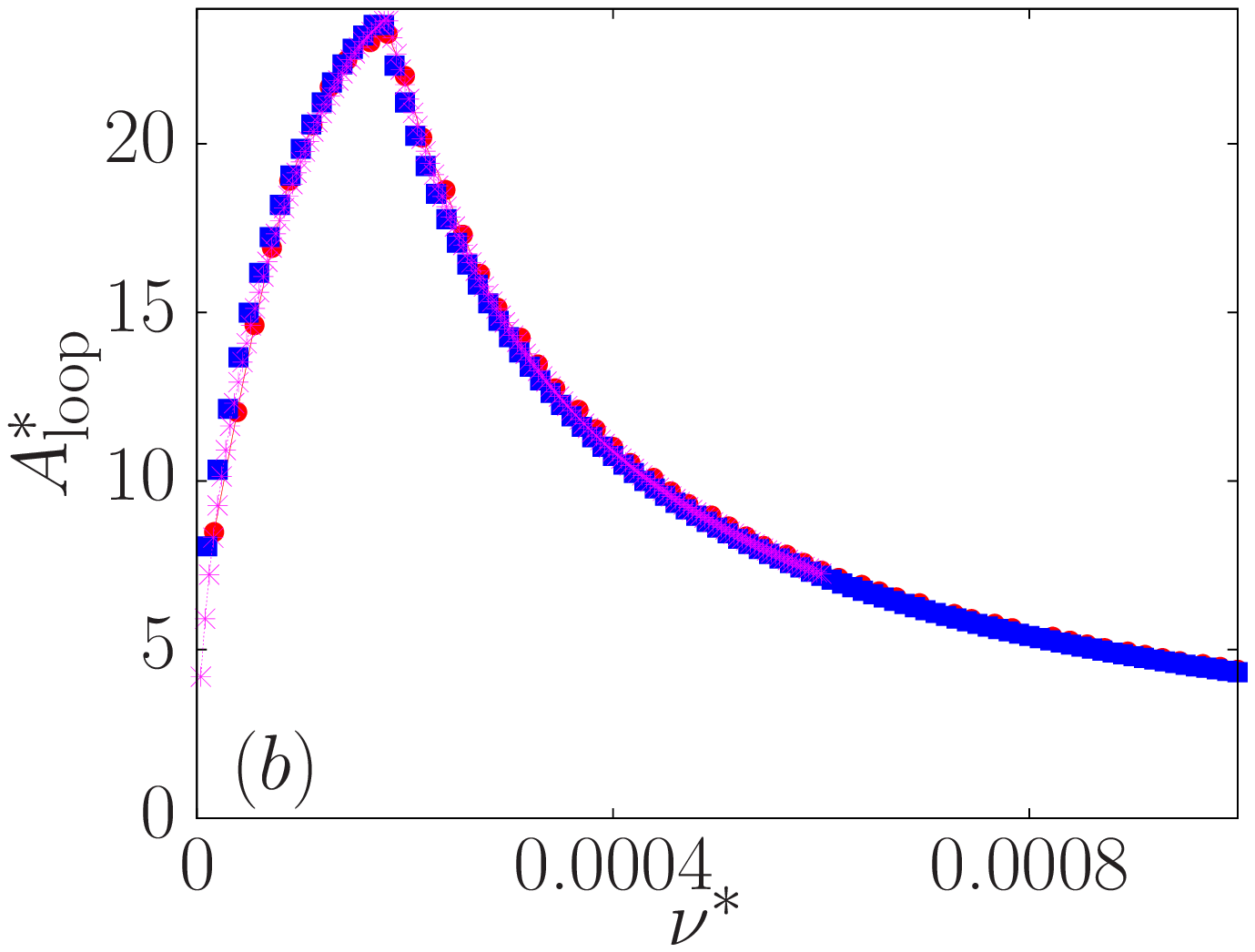}}
        \caption{(Color
  online)   (a) Variation of $A_{\rm loop}$ with $\nu$ for a staircase (LD1) and
sinusoidal force (LD2) using LD simulations, and sinusoidal force using BD simulations.
(b) After rescaling ($A_{\rm loop}^*, \nu^*$), all 
        three curves collapse onto a single master curve, which justifies the use of Eq.\ (3) for
the further understanding.
           }
\label{fig:ana-loop}
\end{figure}
For a quantitative comparison, we plot in Fig.~3(a) for
all three cases the variation of $A_{\rm loop}$ with $\nu$
which exhibits a maximum that corresponds to a critical frequency
\cite{dd}. The variation of the loop area with frequency qualitatively remains similar to the one
seen in 
Ref.\ \cite{kumar}.  Moreover, by rescaling the frequency $\nu^*=a\nu$ for the
over-damped case and rescaling the area $A^*_{\rm loop}=bA_{\rm loop}$ for the staircase 
results, all three curves collapse onto a single master curve [Fig.~3(b)].  This conveys that the
qualitative features and associated scaling  will not change, if one performs simulations in the
over-damped limit with sinusoidal force. Therefore, Eq.\ \ref{lang} 
can now be put in the following form: 
\begin{equation}
\frac{d{x}}{dt} = \dot{x} = -\tilde{\epsilon} \frac{{x}}{|x|} + \tilde{F} \sin(\omega t), 
\label{tanh}
\end{equation}
where  $\tilde{\epsilon}=\epsilon/\zeta $ and $\tilde{F}=F/\zeta$ are 
re-scaled values of $\epsilon$ and $F$, respectively. The area of the hysteresis loop scales as 
\begin{align}
 A_{\rm loop} &\simeq (\tilde{F}-\tilde{\epsilon})^\alpha \omega^\beta L^\gamma 
  &\simeq \left(\frac{F-F_c}{\zeta}\right)^\alpha \omega^\beta L^\gamma 
\, 
,\label{scaling_loop}
\end{align}

\noindent where $F_c \equiv \epsilon$ is the critical force for the unzipping and $\gamma$ is the
exponent associated with length. 
Equation (4) implies that the scaling is independent of $\zeta$ \cite{text2}.

Even under this simplified description, because of the singularity at $x = 0$, the analytical
solution 
of Eq.\ (3) is not easy. However, imposing physical boundary conditions (discussed below), 
its solution  has the form
\begin{equation}
x(t) =
\begin{cases}
c_1-\tilde{\epsilon} t - \frac{\tilde{F}}{\omega} \cos(\omega t), & \text{if 
}x>0  \\
c_2+\tilde{\epsilon} t - \frac{\tilde{F}}{\omega} \cos(\omega t), & \text{if 
}x<0
\end{cases}
\label{eq:anal_solution}
\end{equation}
\begin{figure}[!t]
        \centering
        \subfigure{\includegraphics[scale=0.29]{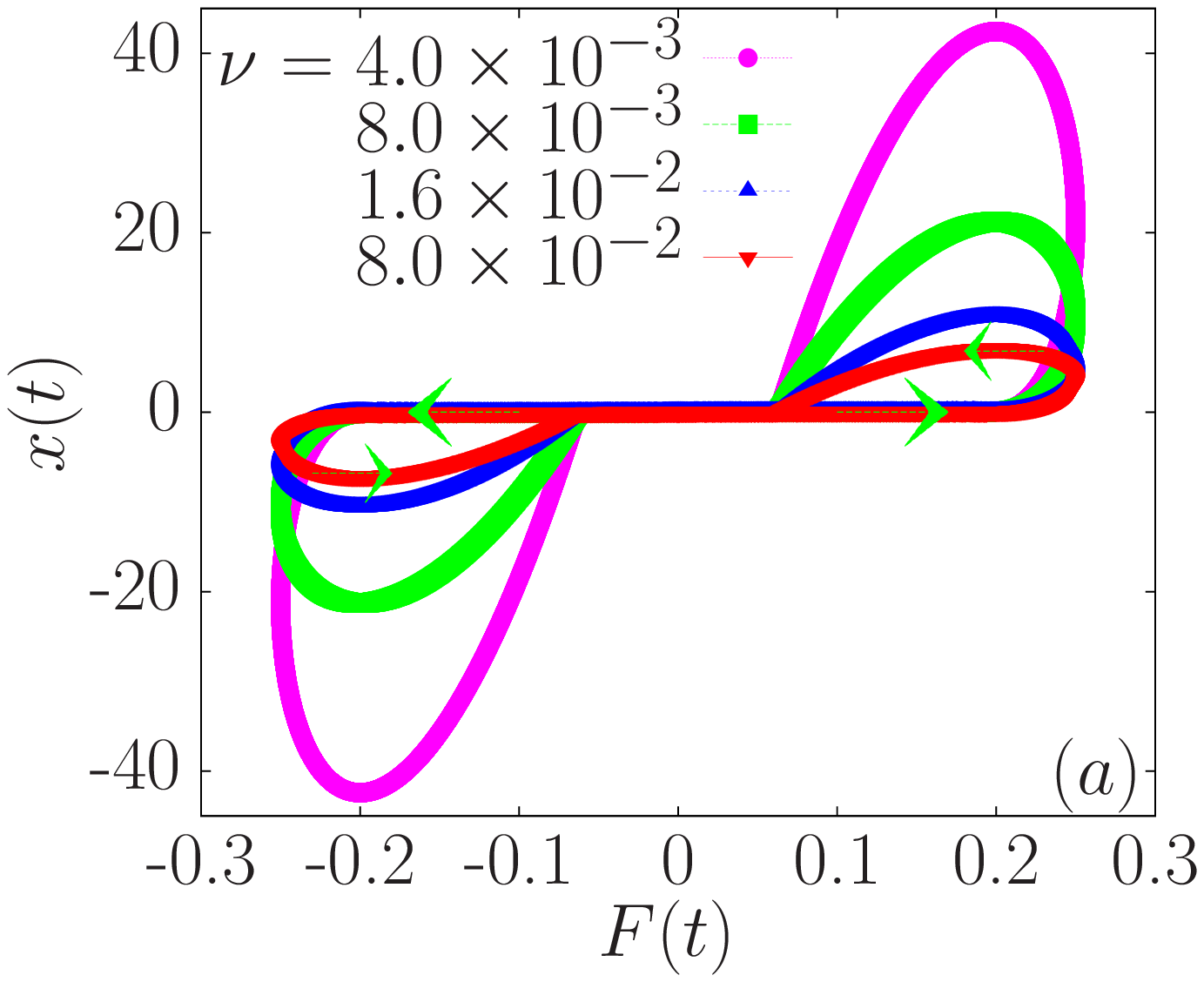}}
        \subfigure{\includegraphics[scale=0.29]{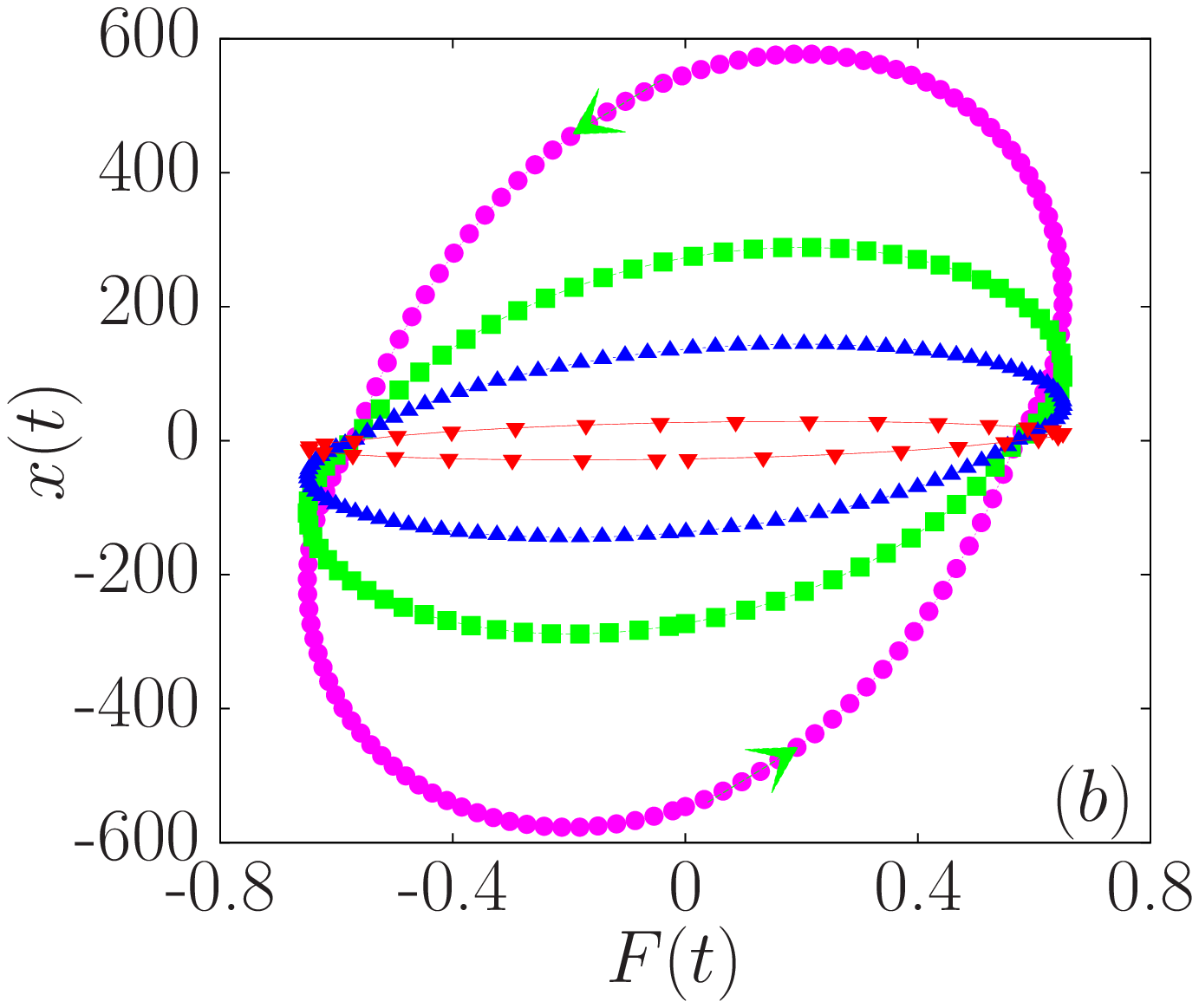}}
        \subfigure{\includegraphics[scale=0.29]{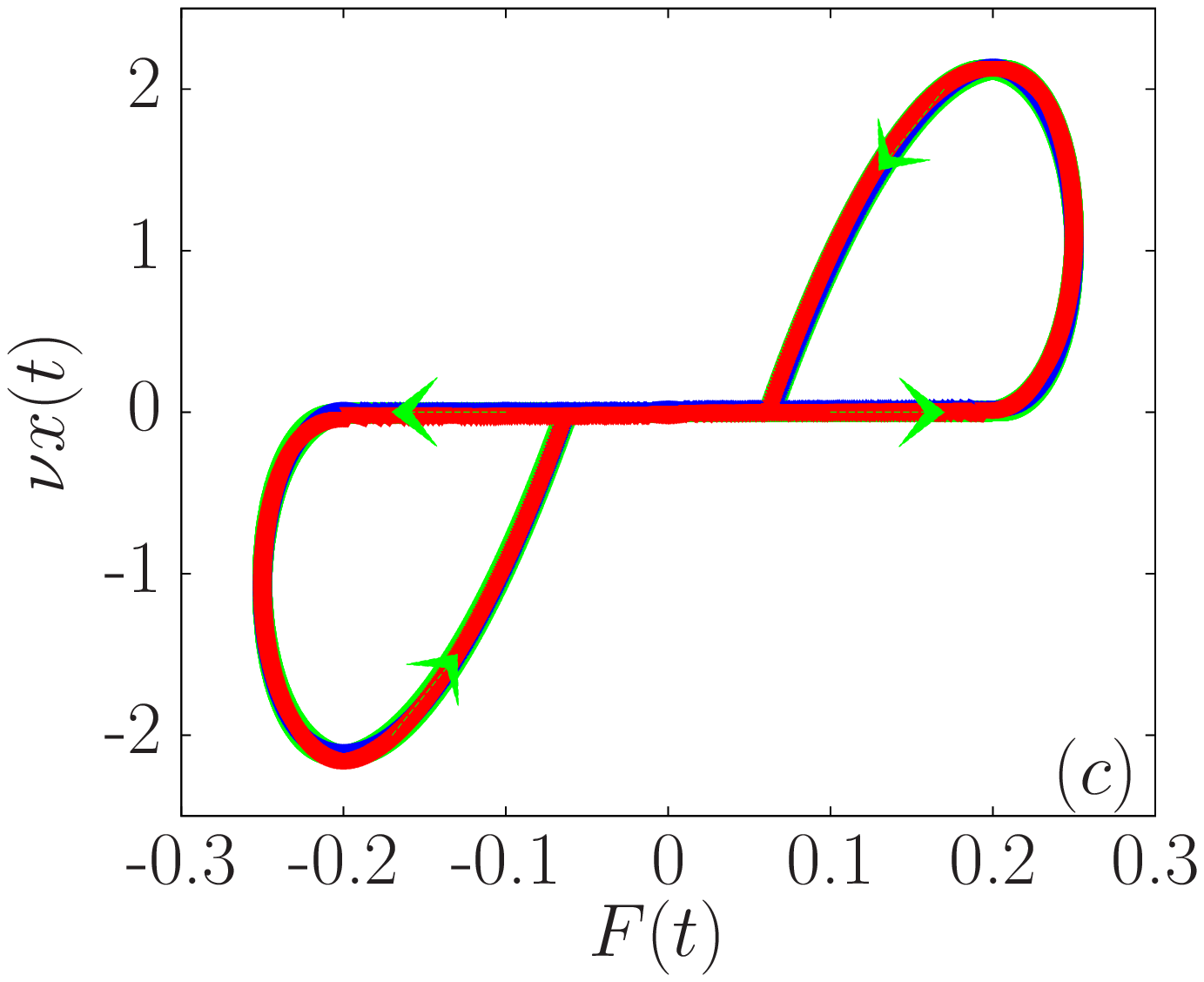}}
        \subfigure{\includegraphics[scale=0.29]{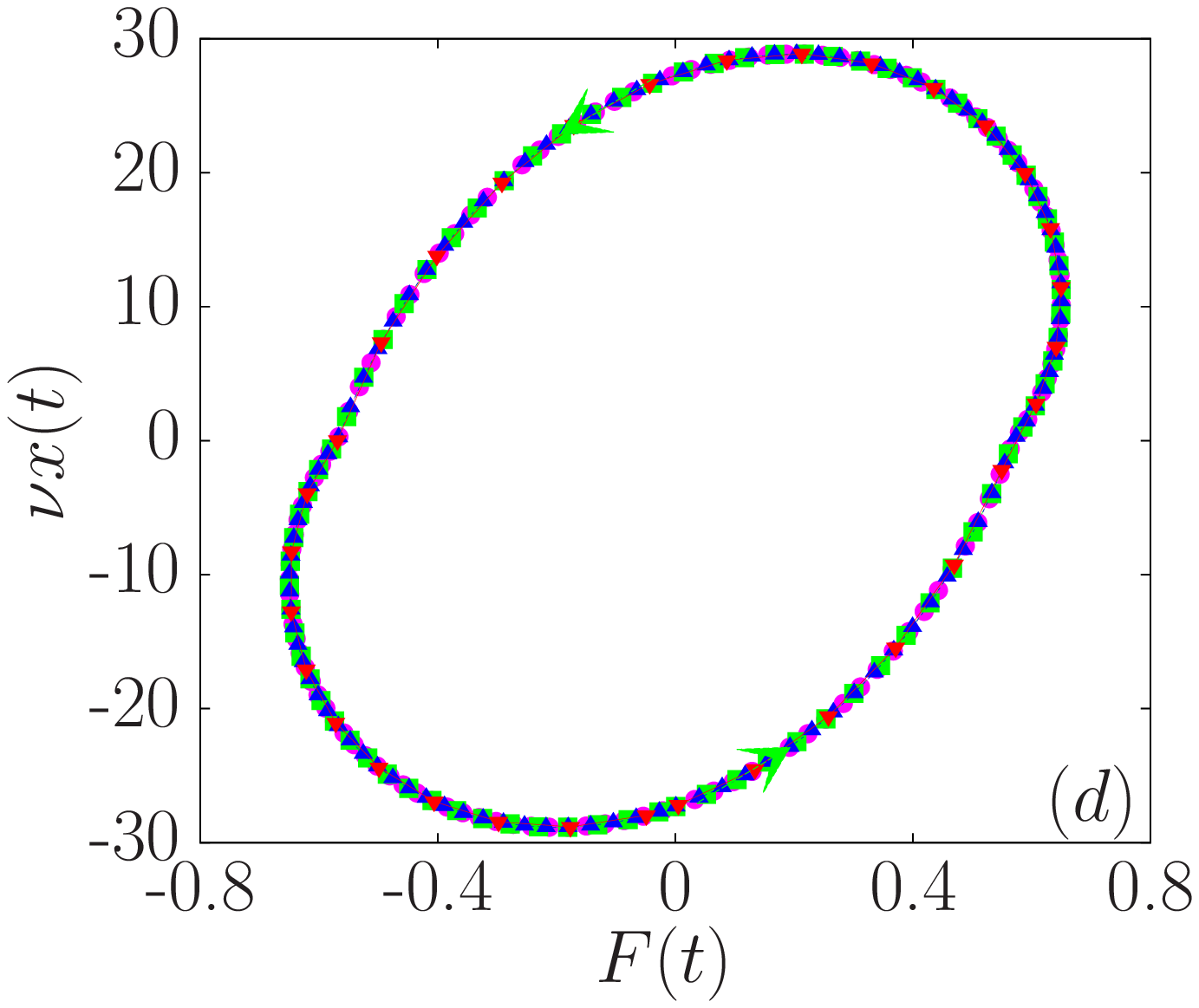}}
        \subfigure{\includegraphics[scale=0.29]{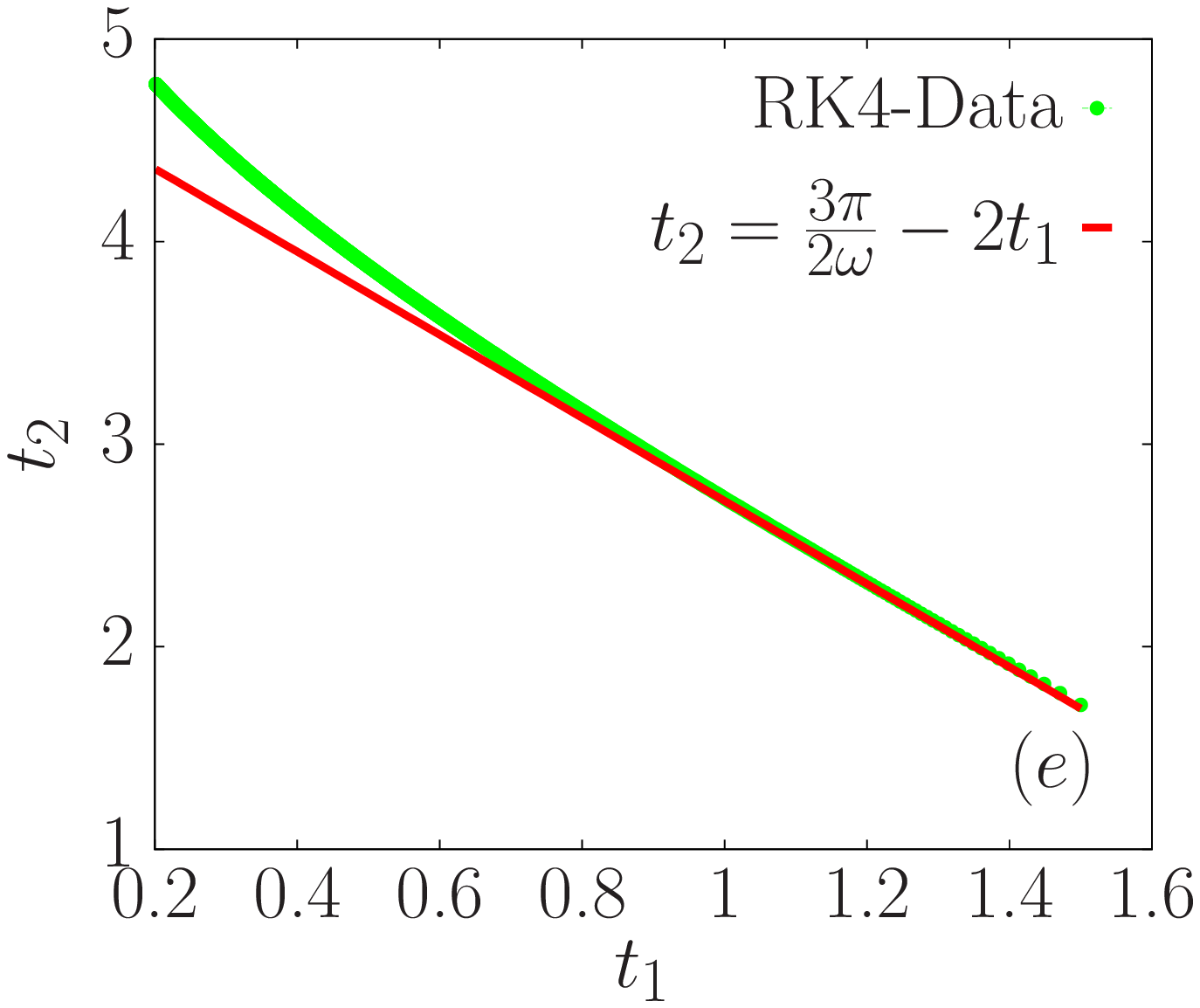}}
	\subfigure{\includegraphics[scale=0.29]{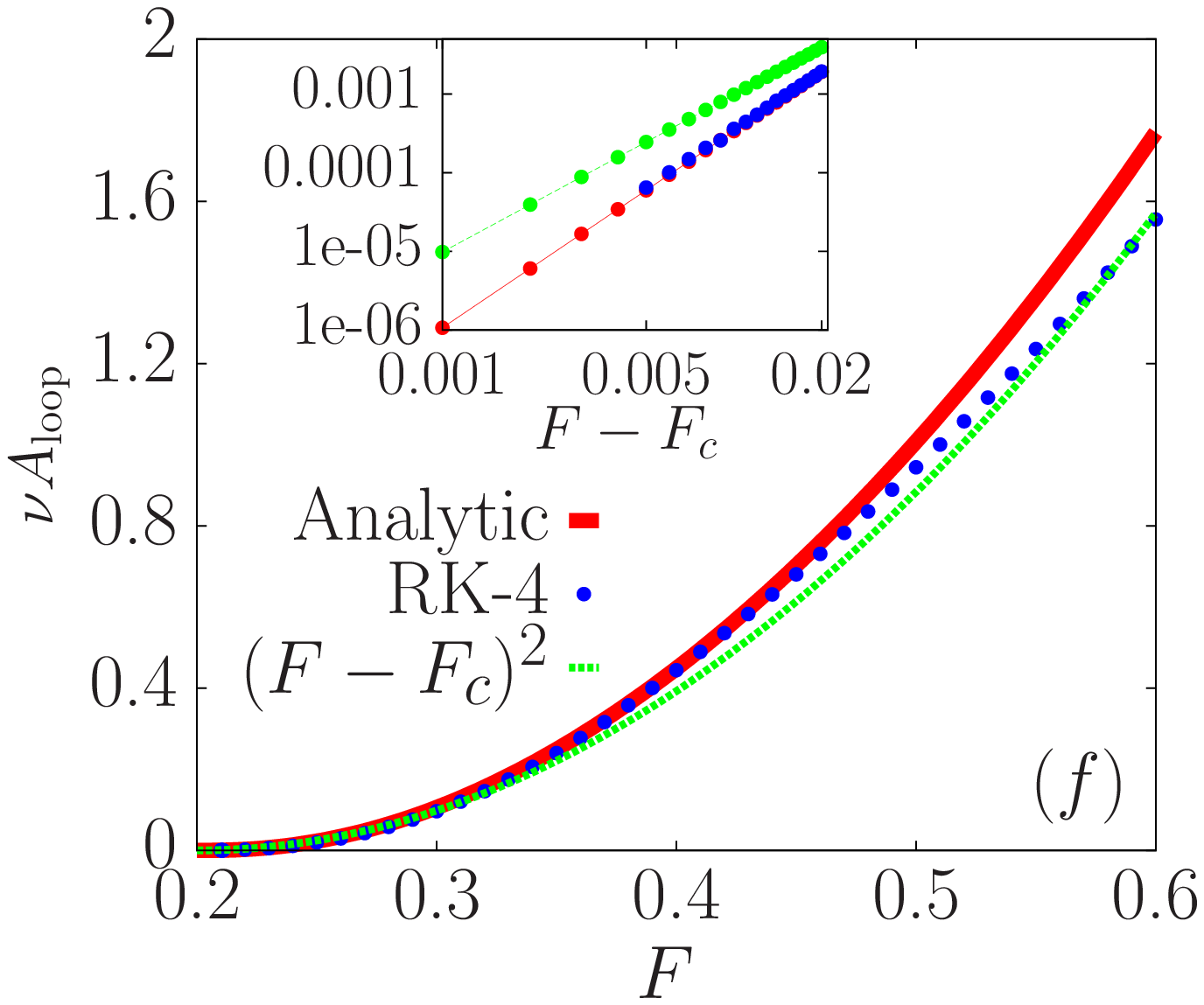}}
        \caption{(Color
  online)  (a), (b) show the $F-x$ curves at different frequencies obtained from Eq.\ (3) for $F =
0.25$ 
        (low) and $0.65$ (high) amplitudes, respectively.  (c), (d) show the collapse of the data 
        in  (a), (b) onto a single hysteresis curve. (e) Comparison of numerical (RK4) values of
$t_2$ as a function of
         $t_1$ with the approximate value for $\omega =1$. (f) shows the comparison of $\nu A_{\rm
loop}-F$ curves 
        obtained from different approaches [Eqs.\ (3), (4), and (6)]. The inset of
Fig.~4(f) is a log-log plot showing that the numerical data (RK4) agree with the expansion of Eq.~(6) in the limit $y \to 1$ $(F
\to F_c)$, i.e., Eq.\ (4) scales with $\alpha= 2.5$ instead of $2$.  
        }
\label{fig:criticalLL}
\end{figure}
\begin{figure*}[t]
        \centering
        \subfigure{\includegraphics[scale=0.29]{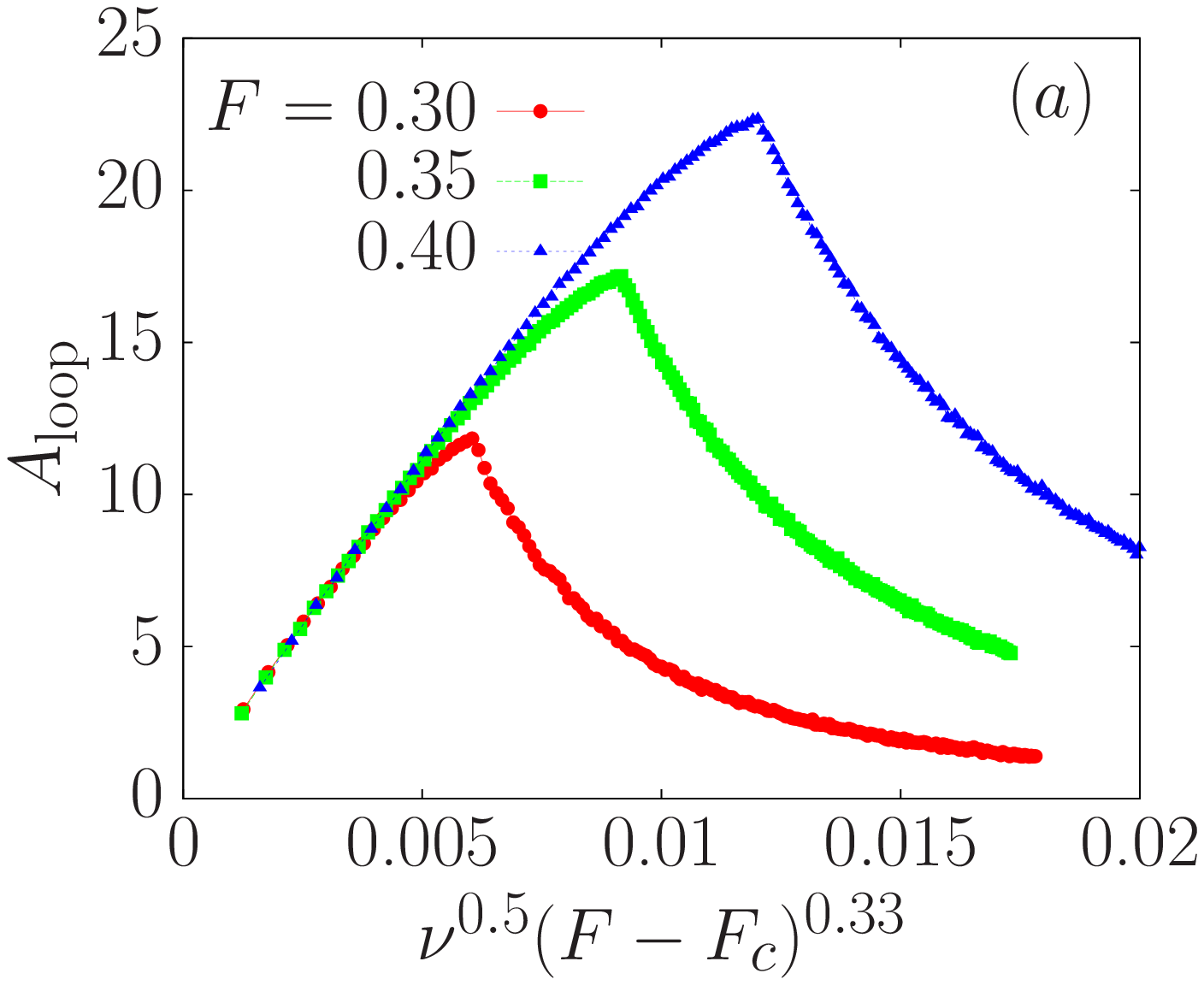}}
        \subfigure{\includegraphics[scale=0.29]{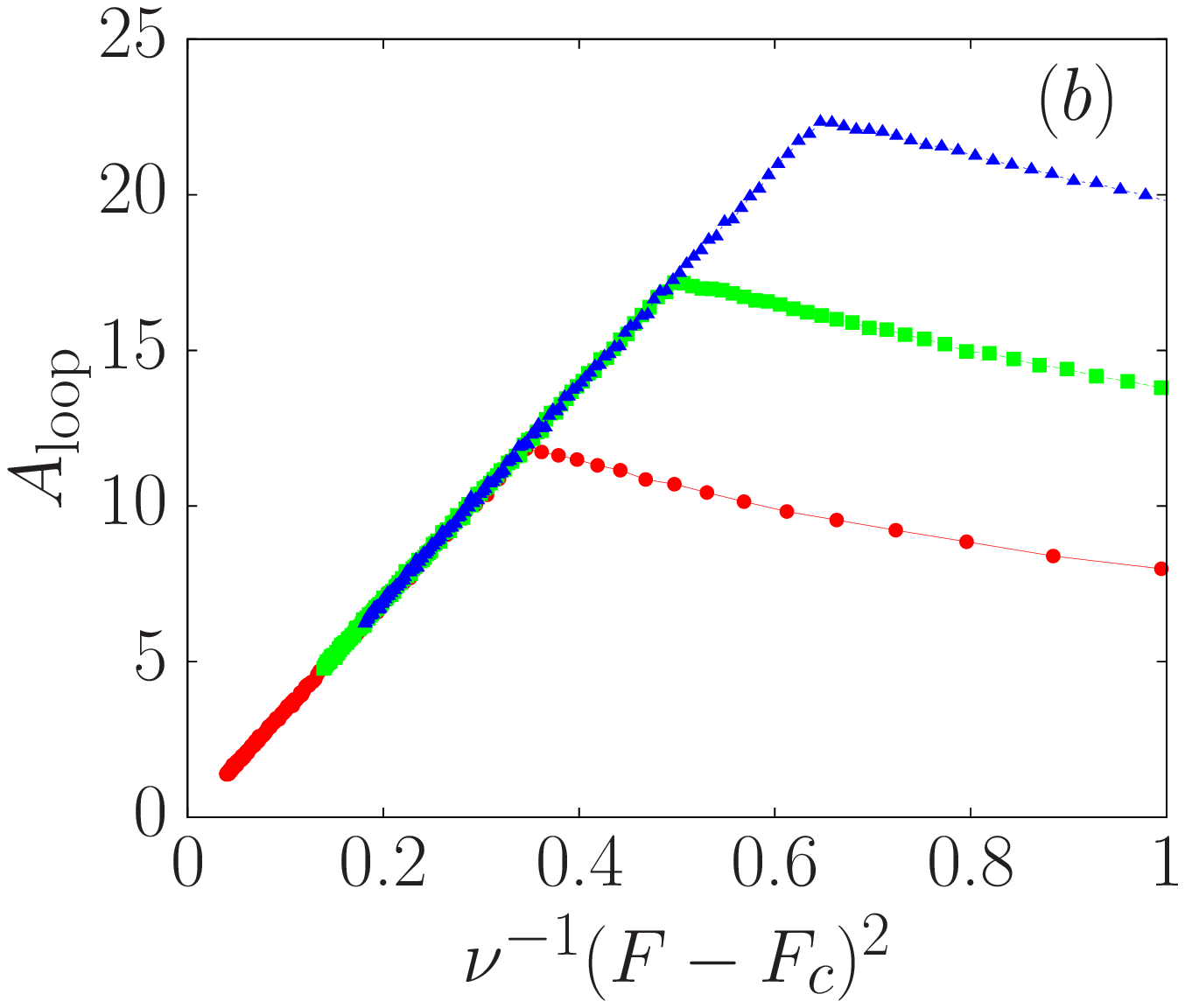}}
        \subfigure{\includegraphics[scale=0.29]{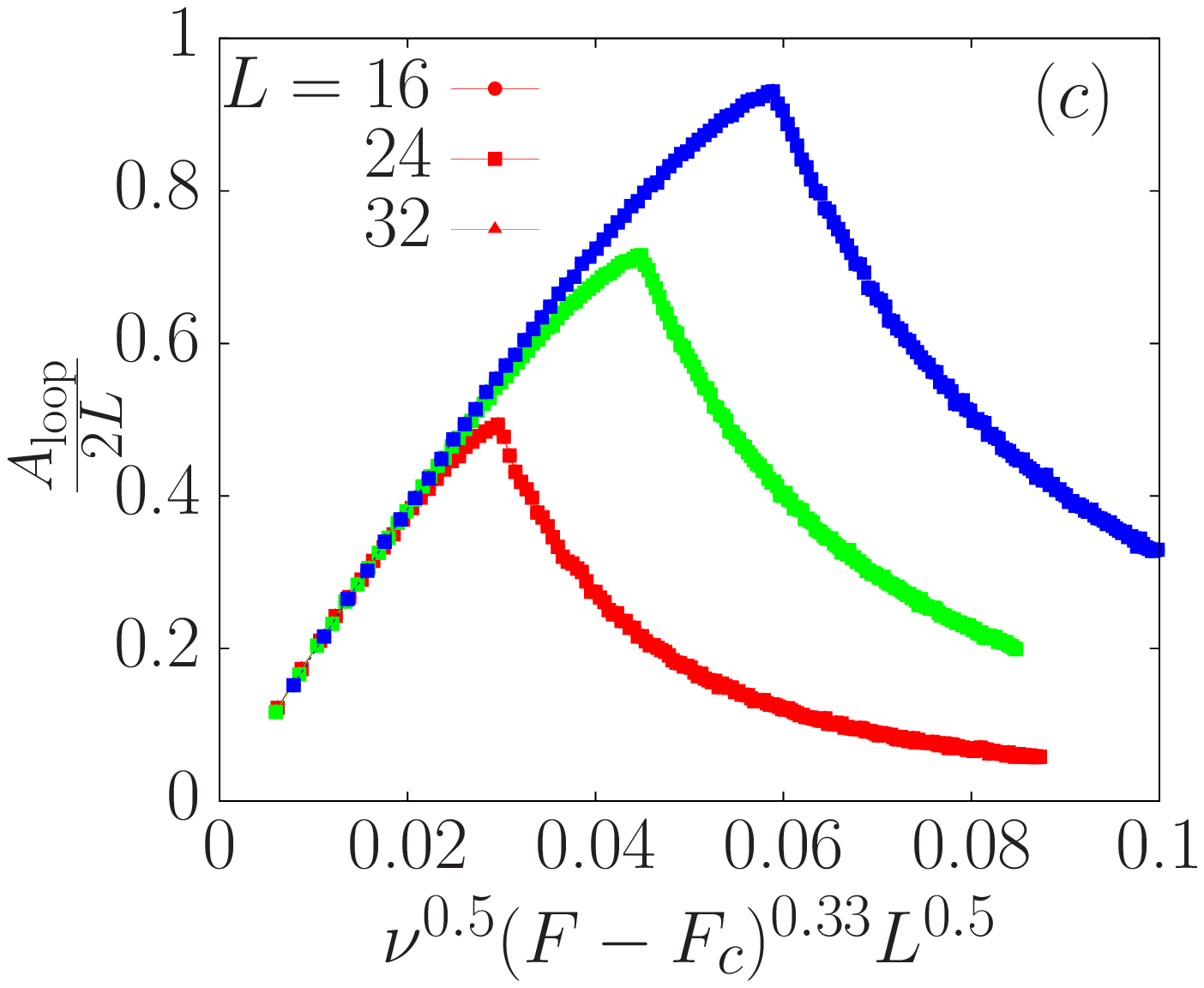}}
        \subfigure{\includegraphics[scale=0.29]{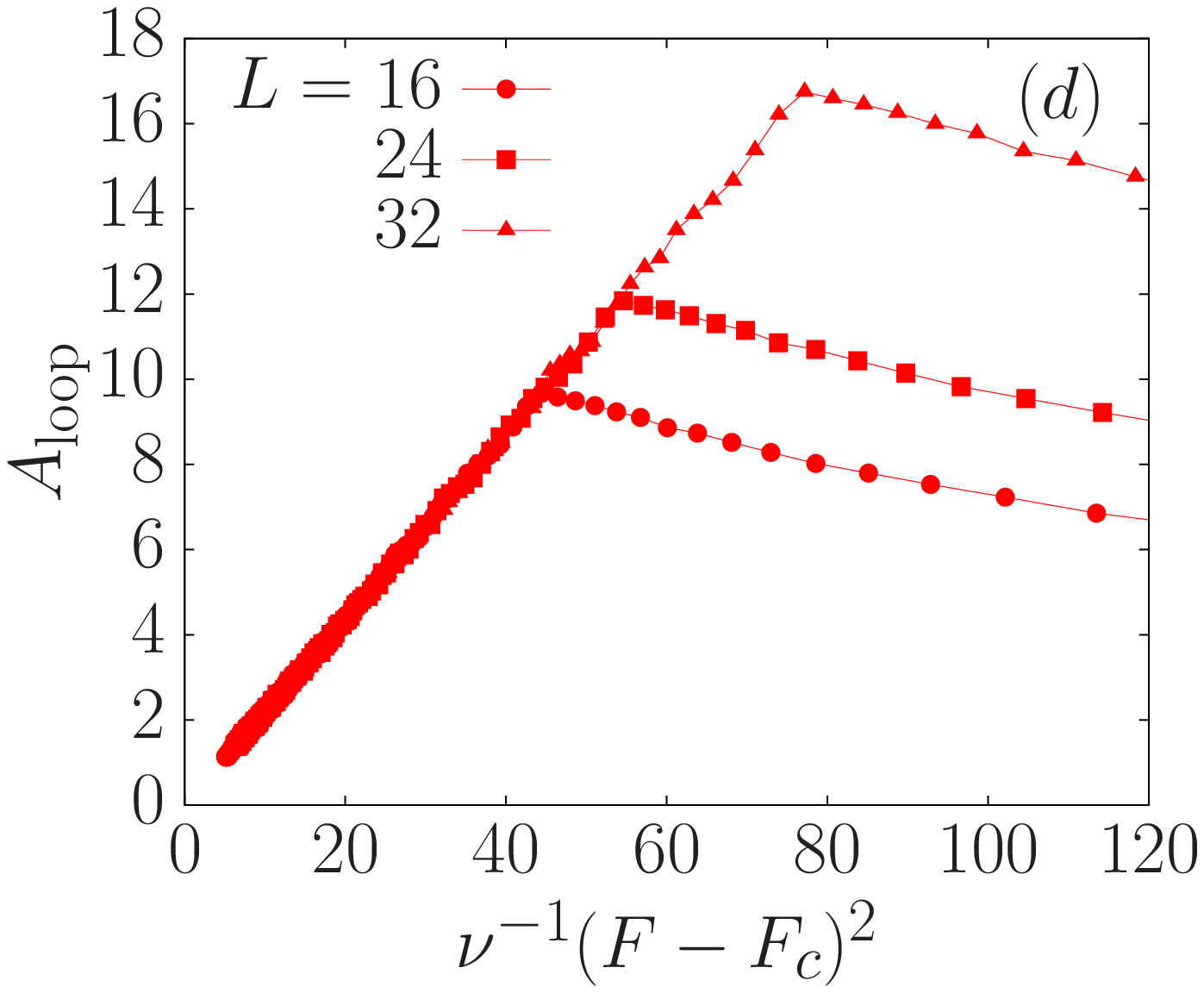}
}
        \caption{(Color
  online) Scaling of $A_{\rm loop}$ with respect to (a) $\nu^{0.5}(F-F_c)^{0.33}$ in the
low-frequency regime and 
        (b) $\nu^{-1}(F-F_c)^{2}$ in the high-frequency regime for a fixed length $L=24$.  (c)
Scaling of 
        $\frac{ A_{\rm loop}}{2L}$ with respect to $\nu^{0.5}(F-F_c)^{0.33}L^{0.5}$ in the
low-frequency regime. Each color represents here three different values of $L$. (d) Scaling
of $A_{\rm loop}$ with respect to  $\nu^{-1}(F-F_c)^{2}$ for a particular value of $F$ $(=0.3)$, demonstrating the length independence in the high-frequency limit.}
\label{fig:scaling}
\end{figure*}

\noindent which (due to relaxing the constraint $|x| \le x_{\rm max} = 2L$) corresponds to the 
asymptotic limit $L \to \infty$. Here, $c_1$ and $c_2$ are the constants of integration, which
can be evaluated by substituting $x(t) = 0$. 
Let us assume that at time $t = 0$, the DNA is in the fully zipped state. As time $t$ elapses,
the magnitude of the applied force increases. If the magnitude of the applied 
force is less than the equilibrium critical force $F_c$, DNA remains in the zipped state 
and as a consequence the velocity of the bead, where the force is applied, 
remains zero. After a certain time, $t_1 = \frac{1}{\omega} \sin^{-1}(\tilde{\epsilon}/\tilde{F})$,
the applied force exceeds $F_c$, and the bead follows the force. After time  $t^{\prime} =
\frac{\pi}{2\omega}$,
 the magnitude of the applied force acquires its maximum value. Thus time 
needed to reach maximum velocity from zero is $\frac{\pi}{2\omega} -t_1$. With further increase in
time,  
the extension increases until $\dot{x}$ becomes zero, i.e., up to time $t^{''}
=\frac{\pi}{\omega}-t_1$.
 After that the extension approaches towards zero with maximum velocity. If we assume the time
needed 
 ($\frac{\pi}{2\omega}-t_1$) to increase the velocity from zero to maximum remains the same then the
time $t_2$ 
 needed to reach $x(t) =0$  can be approximated as $\frac{3\pi}{2 \omega}-2t_1$. In Fig.~4(e) we
have 
 plotted $t_2$ as a function of $t_1$ obtained numerically [from Eqs.\ (3) or (5)] and from the
  approximate form. A nice agreement can be noticed at low force for the hysteretic state. However,
   these values differ at high force, where DNA always remains in the open state over the cycle.  
   In such case, the values of $t_1$ and $t_2$ shift continuously over cycles until they acquire 
   steady-state values after many cycles, and therefore, one has to resort to their numerical
values.

The lag between 
the applied force and the extension constitutes hysteresis, which is depicted in Figs.~4(a) and
4(b) for low 
($F = 0.25$) and high ($F = 0.65$) amplitudes  of the force at different frequencies, respectively. 
In contrast to a finite chain length $L$, where the area of the hysteresis loop first increases and
then decreases 
(Fig.~2), here, the area of the loop always increases with decreasing frequency. Multiplying
numerator and denominator of Eq.\ (3) by $\omega$, it can be shown that $\nu x(t)$ will be a
constant
implying that all curves of different $\nu$ should collapse onto a single curve. This indeed we see in
Fig.~4(c),~(d). 
In fact, for a given length  $L$, 
there exists a critical  frequency $\nu_c$ such that for  $\nu > \nu_c$, the scaling is $L$
independent and the system 
inhibits the same solution as in the limit $L\rightarrow \infty$ and {\bf {\it vice versa}}. 

The area of hysteresis loop $ A_{\rm loop}$ may be calculated numerically. By symmetry,
$A_{\rm loop}$ will be equal to $2 \int_{t_1}^{t_2} F(t) \dot{x}(t)dt$. Because of the
transcendental nature of Eq.\ (5), 
an analytical  expression for $t_2$ appears to be difficult. The approximate value of $t_2$
discussed above [Fig.~4(e)] leads to
\begin{equation}\label{area_ana}
%{\scriptsize
\footnotesize
A_{\rm loop} =\frac{F^2}{\omega \zeta^2}\left(y(2y-3)(1+2y)\sqrt{1-y^2}+3\cos^{-1}(y)\right),
\end{equation}
where $y = \epsilon /F$. Because of the approximation involved in $t_2$, Eq.\ \ref{area_ana} is
still in an approximate
 form. In Fig.~4(f), we plot $A_{\rm loop}$ as obtained from  Eqs.\ (3), (4), and (6) with $\nu$ 
for
low and high amplitudes of the
applied force. The nice agreement among the scaling proposed, numerical solution and analytical
(approximate) solution 
reconfirms that the system has only one scaling. In the high-force limit, 
one can see from Fig.~4(f) and Eq.\ (6) that $A_{\rm loop} \approx \nu^{-1}F^2$, which is
consistent
with the earlier 
studies \cite{kumar, mishra,kapri1}. However, in the low-force limit  ($y \to 1$), the leading term
of 
the expansion of Eq.\ (6) is $\frac{42}{5}\sqrt{2} (\frac{F-F_c}{F_c})^{2.5}$, which is consistent 
with the numerical results [inset of Fig.~4(f)] obtained here \cite{new}.

Let us now turn to the case of finite length $L$ and analyze the scaling in the low-frequency
regime. For 
this Eq.\ (3) can be solved numerically by fixing $x = x_{\rm max} = 2L$, over which the chain 
cannot be stretched. In 
Fig.~5(a), we show the variation of $A_{\rm loop}$ with  $\nu^{0.5} (F-F_c)^{0.33}$ 
in the low-frequency regime, and in Fig.~5(b) with $\nu^{-1} (F-F_c)^{2}$ in the high-frequency
limit. 
Interestingly, the scaling involved in frequency for low- and high-frequency  regimes  (Fig.~5)
remains here the same compared to the model having enough mesoscopic details. The scaling 
associated
with $F$ here is found to be equal to 0.33 and 2 in the low- and high-frequency regimes,
respectively \cite{text3}. 
Collapse of  $A_{\rm loop}/2L$ onto a single line for all lengths in the low-frequency 
regime  confirms that the area scales with length as $L^{0.5}$ [Fig.~5(c)]. It is also evident from
Fig.~5(d) that 
at high frequency, scaling remains independent of $L$ as predicted by Eq.\ (6) and seen in simulations 
\cite{mishra}.

This paper reports many unexplored aspects of the dynamical transition associated with DNA unzipping 
under an oscillatory force. By successive elimination of  the degrees of freedom and parameters, 
we developed a minimal model which presumably remains a good description for  a wide range of parameters and captures 
the essential physics of the dynamical transition.
These results are in agreement with Refs.\ \cite{kapri} and \cite{kumar} in the high-frequency limit, but
strongly differ from Ref.\ \cite{kapri} in the low-frequency regime. The
analytical solution based on the minimal model provides unequivocal support for the absence of 
the dynamical transition  in the thermodynamic limit. 
Moreover, scaling remains independent of temperature. The most notable outcome of the present study is the 
existence of a new scaling exponent associated with force
in the low-force regime, which has been overlooked in all the previous studies \cite{kumar, mishra, kapri}.
 While the model developed here neglects mesoscopic details, such as, excluded volume effect, spring nature
  of covalent bonds, helical nature of DNA, heterogeneity in the sequence etc., the robustness of the exponents
   suggests that it is not restricted to the study of DNA only, but may be extended to many other periodically driven complex systems \cite{cata, exper}.

We thank D. Dhar, Y. Singh, and P. Schierz for many helpful discussions on the subject. The financial 
assistance from the DST, New Delhi, India, the DFG Sonderforschungsbereich/Transregio 
SFB/TRR 102  {\em Polymers under Multiple Constraints: Restricted and Controlled Molecular 
Order and Mobility\/}, Halle-Leipzig, and the Graduate School under Grant No.\ CDFA-02-07 of the 
Deutsch-Franz\"{o}sische Hochschule (DFH-UFA), Germany, are gratefully acknowledged.


\begin{thebibliography}{99}
\bibitem{albert} B. Alberts, D. Bray, J. Lewis, M. Raff, K. Roberts, and
J. D. Watson, {\it Molecular Biology of the Cell\/} (Garland Publishing, New
York, 1994).
\bibitem{tom} D. Tomkiewicz, N. Nouwen, and A. Driessen, FEBS Lett.
{\bf 581}, 2820 (2007).
\bibitem{patel}  I. Donmez and S. S. Patel, Nucleic Acids Res. {\bf 34}, 4216 
(2006).
\bibitem{jan} M. E. Fairman-Williams and E. Jankowsky, J. Mol. Biol. {\bf 415}, 
819 (2012).
\bibitem{velankar} S. S. Velankar, P. Soultanas, M. S. Dillingham, H. S. Subramanya, and D. B. Wigley, Cell {\bf97}, 
75 (1999).
\bibitem{wang} D. S. Johnson, L. Bai, B. Y. Smith, S. S. Patel, and M. D. Wang, Cell {\bf 129}, 1299 (2007).
\bibitem{basu}  A. Basu, A. J. Schoeffler, J. M. Berger, and Z. Bryant, Nat. Struct. Biol. {\bf 19}, 538 (2012).
\bibitem{fili} N. Fili, 
% {\it et al.\/}, 
G. I. Mashanov, C. P. Toseland, C. Batters, M. Wallace, J. T. P. Yeeles, M. S. Dillingham, M. R. Webb,
and J. E. Molloy, 
Nucleic Acids Res. {\bf 38}, 4448 (2010).
\bibitem{janovjak} P. Szymczak and H. Janovjak, J. Mol. Biol. {\bf 390}, 443 (2009).
\bibitem{Liphardt} J. Liphardt, S. Dumont, S. B. Smith, I. Tinoco, Jr., and
C. Bustamante, Science {\bf 296}, 1832 (2002).
\bibitem{Collin} D. Collin, F. Ritort, C. Jarzynski, S. B. Smith,
I. Tinoco, Jr., and C. Bustamante, Nature {\bf 437}, 231 (2005).
\bibitem{Schlierf} M. Schlierf, F. Berkemeier, and M. Rief,
Biophys. J. {\bf 93}, 3989 (2007).
\bibitem{kumarphys} S. Kumar and M. S. Li, Phys. Rep. {\bf 486}, 1 (2010).
\bibitem{maren} A. K. Chattopadhyay and D. Marenduzzo, Phys. Rev. Lett. {\bf
98}, 088101 (2007).
\bibitem{kapri} R. Kapri, Phys. Rev. E {\bf 86}, 041906 (2012).
\bibitem{kumar} S. Kumar and G. Mishra, Phys. Rev. Lett. {\bf 110}, 258102
(2013).
\bibitem{arxiv1} G. Mishra, P. Sadhukhan, S. M. Bhattacharjee, and
     S. Kumar, Phys. Rev. E {\bf 87}, 022718 (2013).
\bibitem{mishra} R. K. Mishra, G. Mishra, D. Giri, and S. Kumar, J. Chem. Phys.
{\bf 138},
244905 (2013).
\bibitem{kapri1}R. Kapri, Phys. Rev. E {\bf 90}, 062719 (2014).
\bibitem{madan1} M. Rao and R. Pandit, Phys. Rev. B {\bf 43}, 3373 (1991).
\bibitem{madan2} M. Rao, H. R. Krishnamurthy, and R. Pandit, Phys. Rev.
     B {\bf 42}, 856 (1990).
\bibitem{dd} D. Dhar and P. Thomas, J. Phys. A {\bf 25}, 4967 (1992).
\bibitem{bkc} B. K. Chakrabarti and M. Acharyya, Rev. Mod. Phys.
     {\bf 71}, 847 (1999).
\bibitem{Li} G. Mishra, D. Giri, M. S. Li, and S. Kumar, J. Chem. Phys. {\bf 135}, 035102 (2011).
\bibitem{rkm} R. K. Mishra, G. Mishra, M. S. Li, and S. Kumar, Phys. Rev. E {\bf 84}, 032903 (2011)
\bibitem{nath} S. Nath, T. Modi, R. K. Mishra, D. Giri, B. P. Mandal, and S. Kumar J. Chem. Phys.  {\bf 139}, 165101 (2013).
\bibitem{cieplak} M. S. Li and M. Cieplak, Phys. Rev. E {\bf 59}, 970 (1999).
\bibitem{libio} M. S. Li, Biophys. J. {\bf 93}, 2644 (2007).
\bibitem{Allen} M. P. Allen and D. J. Tildesley, {\it Computer Simulations of
     Liquids\/} (Oxford University Press, New York, 1987).
\bibitem{Smith} D. Frenkel and B. Smit, {\it Understanding Molecular Simulation\/}
(Academic Press, London, 2002).
\bibitem{new} R. Kumar, S. Kumar, and W. Janke, 
% {\it to be published.}
to be published.
\bibitem{text0} The energy of the model system in Ref.\ \cite{kumar} involves 
a Lennard-Jones potential, which is computationally expensive.  
\bibitem{text1} To avoid this, one can use $F(t) = F \lvert \sin(\omega t)\lvert$. The 
area of the loop will then remain the same, but the frequency will be half compared to the staircase.
\bibitem{text2} This has also been seen  in simulations, however, an analytical argument 
 was beyond the scope of the model in Ref.\ \cite{kumar}.
%\bibitem{text4}  The value of $t_2$ may be obtained from $C_2 - \epsilon t_2+\frac{F}{\omega}\cos(\omega t_2) = 0$.
\bibitem{text3} We revisited the model discussed in Ref.\ \cite{kumar} and 
analyzed the scaling in the low-frequency limit. We found that the loop area in this case 
also scales as $\nu^{0.5} (F-F_c)^{0.33}$, whereas the authors of Ref.\ \cite{kumar} have used 
$\nu^{0.5} F^{0.5}$ to fit their data.
\bibitem{cata} D. Kumar, S. Ghosh,  and S. Bhattacharya, Phys. Rev. E {\bf 87},
013202 (2013).
\bibitem{exper} R. Berkovich, R. I. Hermans, I. Popa, G. Stirnemann, S. Garcia-Manyes, B. J. Berne,
and J. M.  Fernandez, 
 PNAS {\bf109}, 14416 (2012).

% \bibitem{note1} Though,  base pairing (hydrogen bonding) is short-ranged in
%nature, for a
%comparative study (finite length and in thermodynamic limit) we prefer because
%its analytical
%solution can be derived.
\end{thebibliography}
\end{document}